# UNDERSTANDING TEACHER PERSPECTIVES AND EXPERIENCES AFTER DEPLOYMENT OF AI LITERACY CURRICULUM IN MIDDLE-SCHOOL CLASSROOMS

**Prerna Ravi, Annalisa Broski, Glenda Stump, Hal Abelson, Eric Klopfer, Cynthia Breazeal**

*Massachusetts Institute of Technology (UNITED STATES)*

## Abstract

Artificial Intelligence (AI) and its associated applications are ubiquitous in today's world, making it imperative that students and their teachers understand how it works and the ramifications arising from its usage. In this study, we investigate the experiences of seven teachers following their implementation of modules from the MIT RAICA (Responsible AI for Computational Action) curriculum. Through semi-structured interviews, we investigated their instructional strategies as they engaged with the AI curriculum in their classroom, how their teaching and learning beliefs about AI evolved with the curriculum as well as how those beliefs impacted their implementation of the curriculum. Our analysis suggests that the AI modules not only expanded our teachers' knowledge in the field, but also prompted them to recognize its daily applications and their ethical and societal implications, so that they could better engage with the content they deliver to students. Teachers were able to leverage their own interdisciplinary backgrounds to creatively introduce foundational AI topics to students to maximize engagement and playful learning. Our teachers advocated their need for better external support when navigating technological resources, additional time for preparation given the novelty of the curriculum, more flexibility within curriculum timelines, and additional accommodations for students of determination. Our findings provide valuable insights for enhancing future iterations of AI literacy curricula and teacher professional development (PD) resources.

## 1 INTRODUCTION

There is currently a massive proliferation of digital platforms that utilize Artificial Intelligence (AI) stemming from the rapid growth of data and computational power. While this rapidly evolving digital landscape opens up new avenues for collaboration, it also raises several societal ramifications that need to be addressed. Scholars from recent studies have raised concerns about the extent to which K12 students are aware of the applications as well as consequences of AI in their daily lives [1, 2, 3, 4, 5]. This echoes the need for integration of AI in K12 education curriculum to satisfy the need for computational and scientific literacy among the youth and by extension, their educators. Our work investigates teachers' experiences with the introduction to and deployment of three different modules within a novel, project-based AI curriculum.

The Responsible AI for Social Empowerment and Education Initiative (RAISE) at the Massachusetts Institute of Technology (MIT) seeks to equitably educate a broad spectrum of K12 students, an inclusive workforce, and lifelong learners, cultivating them to act responsibly and actively participate in an increasingly AI-driven society. One avenue for this support is the Responsible AI for Computational Action (RAICA) curriculum, which lays a foundation for AI knowledge at the middle-school level through a series of AI learning modules. RAICA comprises a series of project-based modules that progressively increase not only AI knowledge, but also competence in using AI tools. Curriculum modules focus on ethical thinking as well as computational action [6], the idea that learners should be given the opportunity to create projects that have benefit to themselves or their communities. The modules also use a design thinking framework to lead students through the problem-solving process.

Although it is important to provide teachers and students with an AI curriculum that is enriching and engaging, the task of deploying these in classrooms can be daunting for teachers, especially when they lack familiarity in AI topics. For many teachers, curricular prompting and extensive professional development (PD) sessions are important so that they are not only able to successfully understand and deliver AI curricula in varied classroom contexts, but they are able to do so without having a strong prior

STEM background. Instructors recruited to teach AI often have a strong formal background in teaching, but very little computer science (CS) knowledge, and vice versa [7]. Teacher beliefs like high self-efficacy and teaching efficacy are also correlated with integration of content and curriculum in the case of information technology [8]. Past work from Ayanwale and their colleagues has shown that teachers' confidence in teaching AI is correlated with their intention and readiness to teach it [9]. Previous research has also proposed that the idea of AI for social good may serve as a motivating factor driving teachers to embrace AI literacy curricula [10]. Since teachers who look to teach an AI curriculum come from many different backgrounds, introducing professional development (PD) sessions is crucial for the success of the curriculum. Even CS teachers who have strong technical backgrounds in other CS domains may not have the familiarity with AI to jump into an unknown curriculum without additional instruction.

It is important to consider teachers' evolving perspectives when developing effective teaching concepts and material [11]. Prior K12 AI education research in regular classroom settings has primarily involved researchers rather than K12 teachers as the instructors [1, 12, 13, 14]. These studies hence lack the valuable feedback that can be collected from experts who work directly in classrooms [15]. There is also a dearth of information regarding PD elements needed for teaching AI literacy [16]. Our work seeks to fill these gaps. In our curriculum, teachers were offered ongoing PD during their module implementation, they interacted with RAICA researchers during weekly debrief sessions, and they completed semi-structured interviews post-module implementation. These interactions were designed to increase teacher feedback and input on the curriculum and were an important step in the codesign process.

Our research aims to explore and evaluate the impact of AI literacy curriculum (specifically RAICA) on middle-school educators. We shed light on those facets of the curriculum that enabled our teachers' rapid comprehension of new AI concepts and empowered them to draw from their prior interdisciplinary teaching experiences. This allowed them to effectively and confidently modify and adapt the provided curriculum materials to address the unique requirements of their classrooms. We also outline the tools and resources that empowered our teachers to deliver this content with a strong emphasis on fostering socially responsible AI use. We seek to use their feedback and perspectives to inform design changes in future iterations of the curriculum and teacher PD and identify elements from the current deployment that can be retained to strengthen the future landscape of K12 AI literacy.

## 2 METHODOLOGY

Researchers and designers of the RAICA curriculum periodically examined how teachers implemented the curriculum in classrooms. For the purposes of this study, we focussed on eight recorded interviews from seven teachers with interdisciplinary backgrounds in art, business, and computer science. Each teacher implemented one of three modules from the RAICA curriculum: in one module, students build an interactive art installation for their school, in another, students create a chatbot that could inform, entertain, or persuade members of their community, and in the third module, students design an app to help their community discover new things about their school. The curriculum materials consist of a Teacher's Guide, PowerPoint slides for presentation of content, a Student Workbook, and assessments. A kanban board[1] is also available for use as a project management tool.

Following implementation of the modules, we conducted semi-structured remote interviews with each of the seven teachers. We sought to answer three major research questions:

- RQ1: How do teachers adapt AI literacy modules based on their individual classroom contexts and personal backgrounds?
- RQ2: What are teachers' attitudes and beliefs about AI before, during, and after module implementation?
- RQ3: How do teachers' attitudes and beliefs about teaching AI literacy change after using the RAICA curriculum and associated professional development support?

Our study was approved by the Institutional Review Board at Massachusetts Institute of Technology (MIT). Per IRB protocol, we obtained written consent from all interview participants, detailing the purpose of our study and allowing us to record the interview.

---

[1] https://en.wikipedia.org/wiki/Kanban_board

We created an interview protocol with potential questions that could be asked during the interview. The protocol was designed to gather feedback on various aspects of module implementation. Our goal was to investigate teachers' experiences of engaging with the AI curriculum in their classroom, how their teaching and learning beliefs about AI evolved with the curriculum as well as how those beliefs impacted their implementation of the curriculum. During the interviews, we assessed how well teachers and students comprehended the content and their overall impressions of the module. We also identified any challenges they had to overcome during the course of the module, as well as their understanding of AI concepts. We gathered insights into the assessment methods teachers employed and the pedagogical approaches they used within each module to enhance student learning outcomes. The interview questions also inquired about how the module influenced teachers' approaches to teaching and their perceptions of teaching and learning. Through these questions, we aimed to understand how teachers' knowledge of AI had evolved as a result of their experience with the curriculum. We explored any concerns teachers might have had regarding teaching AI in future classes and how this aligned with their personal views on teaching and learning. We also discussed teachers' approach to planning for teaching the AI curriculum.

We transcribed the interviews to initiate the coding process. We followed a thematic data analysis process, with regular check-ins after every interview to discuss findings and emerging themes [17]. Thereafter, we conducted several rounds of deductive coding on the transcribed data. The codes closely followed the text. We identified three major themes to classify codes into: 1) Attitudes and beliefs towards AI literacy 2) Attitudes and beliefs towards teaching and learning, and 3) Teacher recommendations. We also identified subcategories within these three themes to outline more specific trends in the data.

## 3 RESULTS

We now outline the findings that emerged in our data. We present our results around the three primary themes that emerged in the deductive coding process above: attitudes and beliefs towards AI literacy, attitudes and beliefs towards teaching and learning more generally, and recommendations proposed by teachers.

### 3.1 Attitudes and Beliefs towards AI Literacy

This theme explores how teachers and students perceived AI literacy and its various facets. It also investigates specific practices they followed in classrooms to master the complex AI concepts embedded into the curriculum.

*3.1.1 AI Learning*

*3.1.1.1 For Teachers:*

One of the primary goals of our curriculum was to provide teachers with the foundation needed to understand fundamental AI concepts and ideas, so that they could ultimately relay this understanding to diverse student groups in inclusive learning environments. Additionally, teachers need to feel empowered to co-design, implement, and integrate AI curriculum into classrooms with the emphasis on socially responsible use of AI. During our interviews, we investigated teachers' perspectives on learning novel AI concepts. Teachers observed that their engagement with our curriculum facilitated a valuable exploration of AI principles. The pedagogical instruction and lessons within the curriculum played a pivotal role in elucidating the foundational aspects of AI and its broader implications. These experiences led to a transformative shift in their initial perception of AI, which had primarily been influenced by depictions in science fiction films. Their enthusiasm for acquiring this new knowledge stemmed from its practicality in understanding the underlying workings of ubiquitous AI-powered tools employed in their daily lives. For example, teachers engaged in discussions regarding biases inherent in AI systems. They were able to articulate the mechanisms through which bias propagates through datasets in light of emerging AI technologies. They also recognized how their personal experiences and beliefs shaped their students' engagement with AI and their own comprehension of the field. They were hence able to view these from a positive lens:

> "AI can help solve people's problems, but it can also bring harm to a community depending on the group who is making it… but through this curriculum, I feel like we can see these through a positive lens."

Some teachers were also ambivalent about using AI tools: they recognized its dual potential to yield both benefits and harm. These conversations demonstrated their growing awareness of some ethical dimensions of AI. Teachers also valued the peer support and collaboration networks established amongst their fellow educators using the curriculum. It allowed them to exchange and draw inspiration from each other's instructional methods.

However, for all our teachers, AI education exemplified the challenges usually faced during integration of complex unfamiliar technology in classrooms. Teachers ran into technical issues when acquainting themselves with AI platforms like Scratch[2] and Teachable Machines[3]. They advocated the need for one-on-one code reviews in PD sessions, highlighting the significance of personalized guidance when navigating the complexities of AI tools.

#### 3.1.1.2 For Students

Our teachers also spoke about how their students felt about AI and technology after being exposed to the curriculum. They were particularly impressed by their students' deliberation on, and grasp of some complex, abstract ideas related to AI. Their comments were centered around one of the curriculum modules that tasked students with designing and developing a chatbot. As part of this, students delineated the nature and specifics of ideas generated by chatbots, e.g., Kuki[4] and Hector[5]. Both teachers and students gained a comprehensive understanding of the positive roles that chatbots play in informing, persuading, and entertaining users. Students evaluated the chatbot's conversational abilities, emphasizing its capacity to employ human-like language. During these sessions, the primary focus was to gain insight into the students' conceptions of AI and whether they attributed 'consciousness' to these digital entities based on factors like chatbots' friendliness and informativeness. The effectiveness of the chatbot in sustaining meaningful conversations constituted a vital component of the assessment. Some students articulated a viewpoint suggesting that consciousness entailed self-awareness, encompassing attributes such as independent thought and even physiological processes like breathing. This perspective equated consciousness with a tangible, human-like essence. Other students posited that AI, while distinct from human consciousness, possessed the capacity to acquire knowledge and learn over time:

> *"Some students said if you're conscious, it means that you're thinking for yourself and your breathing...this means AI is not conscious in the same way a human is but it can learn things… that is a part of the curriculum."*

Thus, they perceived AI systems as non-human entities capable of learning, in a manner distinct from human consciousness. This perspective led to contemplations on the programming aspect of AI and its capacity for knowledge acquisition.

### 3.1.2 Teaching about AI

#### 3.1.2.1 Strategies

Our teachers also mentioned being concerned about the resistance they would receive from students when introduced to new, complex AI concepts. They employed several innovative methods to introduce fundamental AI concepts to increase classroom engagement and playful learning. One instructional strategy employed by them involved encouraging students to adopt an AI-oriented mindset: thinking like the AI does. This approach aimed to immerse students in the thought processes inherent to AI, fostering a nuanced understanding of the technology. Teachers also tried to draw connections between the lessons they were learning and teaching to other non-stem fields, which helped transcend conventional subject boundaries. For example, one of the teachers commented on the interdisciplinary approach of the module they implemented, noting that it blurred the lines between art/design and computer science classes:

> *"The class was an art class disguised as a CS class and a CS class disguised as an art class."*

This creative approach enriched student learning experiences by integrating diverse fields of study. Other teachers navigated between different instructional elements. Informal formative assessment played a pivotal role in ascertaining students' comprehension of technology concepts. Acknowledging student apprehensions and skepticism towards AI, teachers struck a balance by providing tangible examples. This approach effectively mitigated students' concerns and facilitated a smoother learning process in classrooms. Teachers also prioritised scaffolding instruction and support for students, since they deemed it necessary to guide students through demystifying complex AI concepts.

---

[2] https://scratch.mit.edu/

[3] https://teachablemachine.withgoogle.com/

[4] https://chat.kuki.ai/

[5] https://ochatbot.com/hector/

### 3.1.2.2 Ethical Considerations in AI

The multifaceted nature of AI's impact was a central theme in the curriculum. Teachers incorporated examples into their pedagogy that exposed students to the dichotomy of using AI tools that were capable of both transformative benefits and causing potential harm. This holistic perspective empowered students to critically evaluate the implications of AI applications across various domains, which aligned well with teachers' own perspectives on responsible use of AI (that they had gained from learning about AI through the PD sessions). When guiding students in the creation of innovative AI-driven solutions for everyday challenges, teachers emphasized the significance of considering a broad spectrum of users. They focussed on three major domains of ethical considerations–awareness and consideration of stakeholders, intended and unintended impacts, and industry-aligned design values. These gave students the opportunity to retrieve and use prior learning. These activities also gave authentic avenues to consider the societal ramifications of new information that students were learning.

Teachers encouraged students to investigate the pivotal role of biases and personal experiences in shaping AI interactions. Students explored how inherent biases and stereotypes, coupled with their own unique life experiences, could influence AI's responses and behaviors. This heightened awareness underscored the ethical imperative of mitigating bias and promoting inclusivity in AI design, as supported by the curriculum themes of ethical thinking (ET), design for inclusivity, and equity in AI education.

### 3.1.2.3 Positive Outcomes and Benefits of AI Education

Teachers articulated several benefits from engaging with the AI curriculum, that reflected not only in their eagerness to deliver lessons but also in the responses they elicited from students. Teachers placed a significant emphasis on harnessing technology for positive social impact. This focus resonated with teachers and students alike, as they recognized the potential of AI to bring about meaningful changes in society. As part of the curriculum in one module implementation, teachers brought in AI experts to talk to students about the field and their own career trajectories, and to answer any questions they had. This real-time interaction between students and AI professionals proved to be a powerful aspect of the program. The direct engagement piqued students' interests by providing them with insights into the diverse array of AI careers and the multifaceted skills (like creativity and problem-solving) required in the field, extending beyond computer science programming and mathematics:

> *"The curriculum was really helpful in piquing their interest about what their possibilities for careers are… one of the students mentioned that you don't necessarily have to be a computer programmer to get into the field, which I loved.. I thought that was important and we talked about the different types of skills you need… you can have creativity, you can be a good problem solver. I thought it was really powerful that they could see themselves in that career path in the future."*

These conversations also shattered gender stereotypes surrounding AI careers, fostering a sense of inclusivity and encouraging students to envision themselves pursuing these opportunities regardless of their social and economic backgrounds and identities. They gained confidence in their abilities to contribute to the field of AI, not just as users, but active developers of AI systems. Discussions with these experts not only broadened students' horizons but also provided teachers with valuable insights into the real-world applications of AI and their underlying mechanisms. This deepened insight enabled them to impart knowledge with greater confidence, which was in turn instrumental in kindling students' enthusiasm and curiosity for AI. Additionally, the curriculum shed light on the vital role of AI in the workforce, making students aware of its relevance and applicability in the professional world. This awareness served to underscore the importance of AI literacy and its integration into classrooms.

### 3.1.2.4 Integration of AI Literacy into School Curriculum

Our teachers advocated the need for integrating AI education into existing school curricula. They spoke about devising numerous strategies to stimulate the interest of various educational stakeholders to enable this change. One of the teachers used AI careers as a significant point of interest for students, recognizing the inherent allure of the dynamic field. They also placed a strong emphasis on democratizing access to AI education, promoting creativity in AI endeavors, and highlighting practical applications of AI. Their commitment to making AI accessible to all students, regardless of their prior knowledge, aligned with the goal of achieving AI for all, fostering inclusivity and diversity in AI education.

Teachers also firmly believed that AI education played a pivotal role in shaping students' lives in multifaceted ways. This encompassed not only technical skills but also ethical thinking, responsibility, and inclusivity. They championed the idea that students should create projects that hold genuine meaning, prompting them to think critically about the technology they employ and its potential consequences. Additionally, teachers underscored the importance of design thinking, empathy, and community-focused AI projects. They viewed these aspects as essential components of AI literacy, encouraging students to consider the broader impact of their AI creations on society. In their pedagogical approach, they utilized AI tools such as Kuki as a springboard for AI-related conversations. This strategy enabled students to delve into AI topics from various angles, fostering a deeper understanding and appreciation of the field.

*3.1.2.5 Challenges*

Integrating AI curriculum into classrooms presented a range of challenges for teachers, as evidenced by their experiences and feedback, which encompassed technical issues, varied levels of student knowledge, the need for improved teaching materials, and concerns about potential harm associated with AI. Several teachers found themselves dropping AI components of projects due to limited time and technical issues, emphasizing the importance of extensive teacher training and support to prepare educators for the challenges and opportunities presented by future iterations of the curriculum. In section 3.1.1.1, we spoke about the technical difficulties that teachers ran into when using new AI tools. Frustrations stemming from these difficulties underscored the need for robust technical support and resources to facilitate a seamless AI learning and teaching experience amongst teachers. These problems compounded when students came in with diverse levels of prior technical knowledge. These disparities necessitated the need for tailored approaches and one-on-one code reviews to accommodate individual learning needs. Teachers also discussed refining the AI tools presented to them and their students to ensure that they were user-friendly and effective. The challenges in grasping concepts like Teachable Machines and the recognition of potential harm associated with AI highlighted the significance of comprehensive training and pedagogical support. Teachers acknowledged how their individual teaching methods impacted student learning outcomes in very different ways. It also brought out the need for adaptable resources that could cater to the evolving needs of both educators and students.

## 3.2 Attitudes and Beliefs towards Teaching and Learning

This theme covers how teaching and learning attitudes (more generally) evolved with the curriculum and how those beliefs influenced its implementation in classrooms.

*3.2.1 Teaching Approaches and Strategies*

Teachers employed a variety of instructional strategies when implementing AI curriculum in the classroom, drawing from their prior experiences and insights, thus reflecting their overall attitudes and beliefs about teaching and learning. They encouraged students to engage in brainstorming sessions to foster creativity and independence. They also drew on real-world examples to supplement theoretical lessons in classrooms. Teachers showed a strong preference for visual aids for enhancing student understanding. They also encouraged flexibility and diverse self-expression that enabled students to explore AI concepts in novel ways. There were instances when teachers used structured approaches to delivering content, often involving the use of PowerPoint presentations, since it helped guide students systematically through the material. Teachers simplified complex topics by breaking down lessons into smaller chunks and coupled them with student discussions to promote effective communication and presentation skills. Additionally, prototyping and building systems physically played a crucial role in enhancing students' comprehension and engagement by allowing them to apply their learnings to real-world settings.

Teachers took special measures to reinforce the notions of community and identity amongst student groups. They allocated time to discuss student perceptions on these ideas:

> "We covered the whole idea of community and identity several times and reminded students of it…these conversations helped them come up with some fantastic ideas."

Tools like kanban boards played a crucial role in instilling a sense of community in the classroom. Group prototyping sessions served as another avenue for fostering community-oriented thinking. These sessions allowed students to exercise empathy as they brainstormed novel concepts and identified stakeholders for their final projects, simultaneously acknowledging the influence of their own personal experiences in shaping these ideas.

*3.2.2 Assessment and Evaluation*

Teachers employed diverse assessment tools to gauge students' progress and comprehension. To streamline assessments, some teachers created straightforward criteria checklists and used desired learning outcomes to evaluate students' performance and understanding of AI concepts. Other teachers focused primarily on project completion and adherence to its predefined criteria. Most of the teachers used the kanban board to evidence students' planning and design thinking process, but only few of them used the more playful assessments that were designed for the curriculum. Teachers also created opportunities for student peer teaching and evaluations to foster a collaborative learning environment and provide students with diverse perspectives and constructive criticism on their work. They encouraged students to link specific tasks associated with their project to the overarching driving question or goal that they had initially started with. This was however challenging to accomplish. Given the open-ended structure inherent in the project-based learning model of the curriculum, students found the driving question to be too vague, leading to its limited utility in guiding project work. Consequently, they directed their efforts towards fulfilling assigned tasks, without engaging in critical reflection on the task's overarching relevance:

> "Once students got into the project, they just wanted to work and there's very little thought in the process or the reasoning behind it… and so you have to really keep pulling them back into that line of thinking… this was an ongoing battle throughout."

In response, teachers consistently posed probing questions to redirect students' focus towards the central driving question throughout the learning process. They also motivated students to engage in reflections to nurture their understanding of AI concepts and ethical considerations.

*3.2.3 Curriculum Flexibility and Student-Centered Learning*

The implementation of our curriculum underscored the importance of adopting student-driven learning approaches that cater to diverse learning preferences and interpretations. Teachers noted the provisions made within the curriculum to incorporate various modalities of student expression, fostering creativity, and amplifying student voices. While some students preferred assimilating and learning new material through text, some others thrived in participatory classroom discussions and valued interactive visualizations:

> "Students had really nice thoughts and they expressed themselves differently… sometimes it was written, sometimes they shared them through discussions out loud… other times through visual representations… they felt comfortable as a result… it goes back to the flexibility of the program where it does not have to be just via writing."

These variations also influenced their capacity to retain freshly acquired information, with certain students opting to meticulously transcribe their learnings for future reference and recall. The adaptability of the curriculum also empowered teachers to draw upon their distinct pedagogical backgrounds, including those in non-STEM disciplines such as art and business, to establish real-world connections when elucidating new concepts to students. They seamlessly integrated elements of computational thinking and design thinking, encouraging students to independently explore and generate innovative ideas. They promoted ideas of independent problem-solving through tinkering, experimentation, and troubleshooting errors. They showed students how they could harness the feedback they received from teacher guidance and support to iteratively refine their project designs, thus enhancing their overall efficacy and social impact.

## 3.3 Teacher Recommendations

The eight semi-structured interviews brought to light many recommendations that teachers thought would be useful when integrating the curriculum into their own classrooms. These recommended changes can be broken down into two categories: curriculum and material improvements and teaching and support enhancements.

*3.3.1 Curriculum and Material Improvements*

The teachers made many suggestions regarding improving the curriculum and associated material. The RAICA modules come with a Student Workbook for students to answer questions and document their design thinking process, keeping track of ideas they are formulating when learning about AI. One of our teachers thought that the workbook was too long and required their students to do too much writing, conflicting with their typical hands-on teaching style. Another teacher made a more general request for more opportunities for hands-on experiences for their students.

One teacher modified the curriculum to include more peer feedback during the prototyping stage and felt this to be incredibly useful in their classroom. They suggested that more peer feedback be incorporated into the curriculum:

> *"Within those sessions, we could ideate more [and] solve each other's problems as well."*

Conversely, another teacher wanted the curriculum to dive right into practice, making prototyping with feedback a less crucial step. Two teachers found the use of kanban boards for planning to be *"too difficult"* for their students, abandoning their use in their classrooms, while another made the kanban board a collaborative effort between students to promote engagement with the tool.

Teachers requested more supplemental materials for both themselves and students. One teacher wanted to share more resources with their class, requesting additional examples for students.

Some teachers found that the arrangement of the materials presented challenges. One teacher articulated that a central location with links to all necessary materials would have helped them stay more organised, and that students needed a clear place to curate their ideas and projects.

Although one teacher found the curriculum to be too open-ended, they wanted to stress that teachers provide different directions that their students could take if they felt challenged by the open-endedness. This same teacher also pointed out that they would like to see the curriculum made more accessible to students with differential learning needs, which was seconded by another.

### 3.3.2 Teaching Support

Teachers suggested a plethora of methodologies for enhancing teaching and their feelings of support. Technology was a challenge for some teachers. One teacher had a dedicated tech support person in their classroom and found this to be a valuable tool for their students. Another teacher stated that it was very important for teachers to have more experience with *"real-world AI applications"* to succeed with the deployment of the curriculum.

Teachers wanted to help their students see themselves as part of the future of AI augmented technology, one stating that it is important for teachers to be able to guide their students in making meaningful connections between their projects and their personal lives.

Teachers at times mentioned feeling underprepared to deliver curriculum material and ideas in classrooms. One teacher found the teacher's guides to be *"too vague"* and they wanted more specific directions of what to do in their classroom; another expressed a lack of confidence in addressing students' fears and concerns of AI in a positive manner. A different teacher suggested revisiting past ideas on their own practices of design thinking and empathy to apply to this unfamiliar AI context.

One of our teachers recommended including useful resources to walkthroughs and tutorials with the curriculum for other educators to use when increasing their knowledge base. A second teacher added to this idea by requesting videos to enhance their understanding of AI tools:

> *"I found [video tutorials] super helpful to kind of understand the walkthrough and to help the kids with the technology, because I had done it first."*

One of our teachers even went as far as to suggest that more instruction prior to teaching the modules would be useful, stating that they would have loved an extra month to prepare and understand the content before diving into teaching this material.

## 4 CONCLUSIONS

Taken together, our teachers provided valuable information and feedback necessary to inform future curriculum design changes to this context. We summarize the key takeaways from our study below:

- Designing AI literacy curriculum that is flexible and conducive to varied classroom contexts, including diverse student and teacher backgrounds is critical to ensuring its success and long-term sustainability in the classroom. Since some teachers requested more directions and specifications within our materials to help them navigate novel concepts better, future iterations can provide more examples of hands-on activities (that teachers could organise in classrooms) and scaffolding questions (that they can use to stimulate discussions amongst students). The

- open-ended nature of our curriculum empowered teachers to leverage their unique backgrounds and prior teaching experiences when developing materials to explain new AI concepts. This is an aspect that we would like to preserve in future iterations.

- Teacher support and professional development (PD) sessions should precede the implementation of new AI curricula to ensure that teachers with no prior experience are equipped with the resources needed to comfortably deliver foundational AI knowledge. Some teachers were apprehensive about facilitating challenging conversations about AI and its ethical implications with their students. Preparing teachers for engaging in discussions about fears students have of traversing an unknown technical landscape would require additional PD training and examples on moderating difficult conversations in class. This approach has been utilised in other work resulting in teachers' increased comfort in having conversations about AI ethics as well as using AI tools [18]. The inclusion of such materials in the RAICA curriculum would also likely increase teachers' competence and self-confidence in multiple areas, adding to their productivity and efficacy for students.

- Inclusivity is another key factor to consider when deploying AI curricula. Hands-on and unplugged activities (that aren't reliant on device availability amongst students) can increase the accessibility of the curriculum for classrooms with limited access to technology [19], allowing students to learn something about AI without programming themselves. Additional teaching resources and accommodations for students of determination can also increase the impact of lessons, by catering to a wider audience.

Our work throws light on important considerations when developing an AI literacy curriculum and associated PD. It also underscores the need and benefit of continuous feedback during development and initial deployment. Our interviews help us understand the impact of the curriculum in different classroom environments and encourage us to place a stronger emphasis on accessibility and addressing the varied needs of students in future iterations. Teachers' perspectives are invaluable in achieving these goals.

## ACKNOWLEDGEMENTS

This work was supported by funding from DP World. The funding body played no role in the study design or in the analysis and interpretation of data. We wish to acknowledge and thank our partner schools and their teachers who generously gave of their time to implement our curriculum and provide feedback.